\documentclass[10pt]{article}

\usepackage{amsmath}
\usepackage{amssymb}

\usepackage{graphicx}

\usepackage{color} 

\usepackage[labelfont=bf,labelsep=period,justification=raggedright]{caption} 

\makeatletter
\renewcommand{\@biblabel}[1]{\quad#1.}
\makeatother

\date{}

\begin{document}

\begin{flushleft}
{\Large
\textbf{The Role of Non-native Interactions in the Folding of Knotted Proteins}
}
\\
{\bf
Tatjana \v{S}krbi\'{c}$^{1,2}$, 
Cristian Micheletti$^{3,\ast}$,
Pietro Faccioli$^{4,5,\ast}$ 
}
\\
{\bf 1} ECT* - European Centre for Theoretical Studies in Nuclear Physics and 
Related Areas, Villazzano (Trento), Italy,
{\bf 2} LISC - Interdisciplinary Laboratory for Computational Science, 
Povo (Trento), Italy,
{\bf 3} SISSA - Scuola Internazionale Superiore di Studi Avanzati and CNR-IOM Democritos, Trieste, Italy,
{\bf 4} Dipartimento di Fisica, Universit\`a degli Studi di Trento, Povo (Trento), Italy,
{\bf 5} INFN - Istituto Nazionale di Fisica Nucleare, Gruppo Collegato di Trento, 
Povo (Trento), Italy.
\\
$\ast$ E-mail: michelet@sissa.it; faccioli@science.unitn.it; 
\end{flushleft}

\section*{Abstract}
Stochastic simulations of coarse-grained protein models are used to
investigate the propensity to form knots in early stages of protein folding.
The study is carried out comparatively for two homologous
carbamoyltransferases, a natively-knotted N-acetylornithine carbamoyltransferase (AOTCase)
 and an unknotted ornithine carbamoyltransferase (OTCase).
In addition, two different sets of pairwise amino acid
interactions are considered: one promoting exclusively native
interactions, and the other additionally including non-native
quasi-chemical and electrostatic interactions. With the former model
neither protein show a propensity to form knots.
With the additional non-native interactions, knotting propensity
remains negligible for the
natively-unknotted OTCase while for AOTCase it is much enhanced. Analysis of the
trajectories suggests that the different  entanglement of the two
transcarbamylases follows from the tendency of the C-terminal to
point away from (for OTCase) or approach and eventually thread (for
AOTCase) other regions of partly-folded protein. The
  analysis of the OTCase/AOTCase pair clarifies that natively-knotted
  proteins can spontaneously knot during early folding stages and that
  non-native sequence-dependent interactions are important for promoting and disfavouring
  early knotting events.

\section*{Author Summary}
Knotted proteins provide an ideal ground for examining how amino
acid interactions (which are local) can favor their folding
into a native state of non-trivial topology (which is a global property).
Some of the mechanisms that can aid knot formation
are investigated here by comparing coarse-grained folding
simulations of two enzymes that are structurally similar, and yet have
natively knotted and unknotted states, respectively.  In folding simulations that exclusively promote the formation of native contacts, neither protein forms knots. Strikingly, when sequence-dependent non-native interactions between amino acids are introduced, one observes knotting events but only for the natively-knotted protein. The results support the importance of non-native interactions in favoring or disfavoring knotting events in the early stages of folding.

\section*{Introduction}

A much debated problem in the thermodynamics and kinetics of protein
folding \cite{finkelsteinbook,Creighton} is the process of knot
formation in proteins
~\cite{mansfield1,mansfield2,taylor1,nureki_knot_exp,mtase_knot_exp,taylor2}. The
topological entanglement found in naturally-occurring proteins
presents several differences from that of compact or collapsed flexible polymers~\cite{knot_statistics}.  Firstly, the
overall percentage of knotted native states in the protein data bank (PDB) is lower than for globular flexible chains with the same degree of polymerization~\cite{poly1,poly2,poly3,poly4,poly5,poly6,poly7,poly8, DNApack, taylor3}. In addition, the free energy landscape of proteins, unlike that of homopolymers or random heteropolymers, is sufficiently smooth to ensure that the same knot in the same protein location is observed in folded structures~\cite{MallamJackson2012,Yeates, yeates_review,mallam_review,virnau_review, onuchic2}.

These distinctive features are arguably the consequence of concomitant effects,
including the propensity of polypeptides to form secondary structure elements, which enhance their local order  compared to a typical collapsed polymer chain~\cite{knot_statistics}, as well as functionally-oriented evolutionary mechanisms \cite{virnau_plos,micheletti_plos,ShakhnovichComm}.

To gain insight into such mechanisms as well as to highlight general physico-chemical mechanisms favoring knot formation, an increasing number of experimental and numerical studies of knotted proteins have been carried out \cite{knot_statistics, MallamJackson2012,Yeates, onuchic2, jackson1, jackson2,jackson3,jackson4,jackson5, jackson6,shakhnovich,onuchic1}.
Notably, the latest experimental results have clarified that native knots can form spontaneously and efficiently from an unknotted initial state. In particular, Yeates and coworkers \cite{Yeates} successfully designed a protein which can refold into the natively-knotted structure, albeit much more slowly than an unknotted counterpart. Finally, the recent study of Mallam and Jackson showed that newly translated, knot free, YibK molecules fold spontaneously to the native trefoil-knotted state, and that chaperones can significantly speed up the folding process~\cite{MallamJackson2012}.

Numerical simulations have been a valuable complement to experiments, particularly regarding the characterization of the pathways leading to the self-tying of knotted proteins. In particular, folding simulations based on simplified protein representations and/or force fields, have indicated two main mechanisms leading to knot formation: the threading of one of the termini through a loop~\cite{shakhnovich}, or by slipping a pseudoknot through a loop~\cite{onuchic1}. The two mechanisms are not mutually exclusive, as reported by Sulkowska 
\emph{et al.} \cite{onuchic2} for protein MJ0366 from {\em Methanocaldococcus jannaschii}. Importantly, in coarse-grained (C$_\alpha$-trace) simulations where folding was promoted by exclusively favoring native contacts it was seen that the yield of folding trajectories was low and knots would form at late folding stages. Specifically, for the deeply-knotted protein YibK, only 1-2\% of the trajectories reached the native state and the native knot was formed when about $q\sim80\%$ of the native contacts had already been established. Arguably, the coarse-grained nature of the models contributes to the observed low folding yield. However, even when employing atomistic representations, purely native-centric models tend to favor a rather 
late onset of knotting~\cite{faisca_mc}. For example, a relatively late stage of knot formation $q\sim40\%$ was also estimated from the analysis of the free-energy landscape of 
protein MJ0366, despite being half as long as YibK and with a shallower knot~\cite{onuchic2}.

These findings are aptly complemented by the early observation of Wallin {\em et al.} \cite{shakhnovich}
that the low yield of purely native-centric models can be dramatically enhanced by promoting
the attraction of specific protein regions that are not in contact in
the native state.  In particular, it was seen that the formation of the native knot in
YibK was particularly facilitated by introducing an {\em ad hoc} non-native
attraction between a loop and the C-terminus that, by threading the former, produced
a sizeable population of knots even at early folding stages, $q\sim 20$\% \cite{shakhnovich}.

These results, together with recent experimental ones \cite{MallamJackson2012,Yeates,jackson6}, pose the question of  
whether natively-knotted proteins can form non-trivial entanglements during early folding stages and whether non-native interactions can provide a general mechanisms for such self-tying events. Clarifying such aspects is important to advance the understanding of some of the general mechanisms aiding knot formation. It must however, be borne in mind that there could co-exist independent pathways where knots are established at different stages of the folding process and that active mechanisms, such as interactions with chaperones, could be involved {\em in vivo}~\cite{MallamJackson2012}.
 
\begin{figure}[!ht]
\begin{center}
\includegraphics[width=12cm]{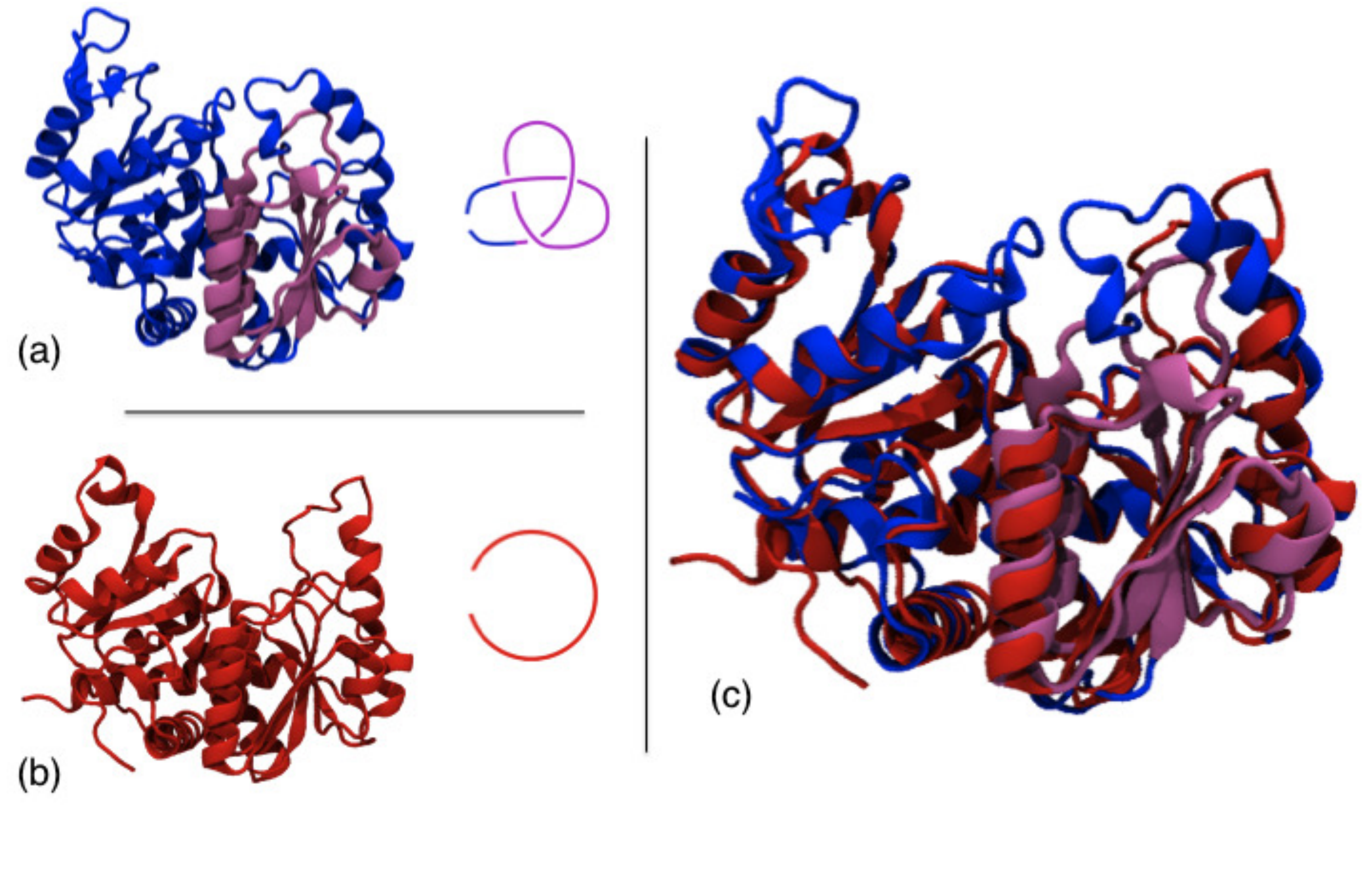}
\end{center}
\caption{{\bf Knotted and unknotted carbamoyltransferases.} 
(a) Cartoon representation of the AOTCase
which is natively-knotted 
in a right-handed trefoil knot (see sketch). The knotted region 
is highlighted in purple. (b) Cartoon representation of the unknotted OTCase.
The MISTRAL structural alignment \cite{mistral} of the knotted AOTCase and unknotted 
OTCase is shown in panel (c).}
\label{Figure_1}
\end{figure}

In this study we address these questions by simulating the early folding process of two transcarbamylase
proteins that are structurally very similar and yet
their native states are differently knotted~\cite{virnau_plos,micheletti_plos}.
Specifically, one of them is a trefoil-knotted N-acetylornithine carbamoyltransferase (AOTCase), while the other
is an unknotted ornithine carbamoyltransferase (OTCase), see Fig.~\ref{Figure_1}. Their evolutionary
relatedness  has posed the interesting question
of understanding the source of their different native topology, with particular regards to the role of specific loop regions whose "virtual" excision or addition alters the native topology \cite{micheletti_plos}.

Here, for both the OTCase and AOTCase, more than a hundred folding simulations are carried
out using a coarse-grained model and two different energy
functions. We first evolve a fully extended configuration and steer it
towards the native state by exclusively promoting native contact
interactions. In a second set of simulations we enrich the energy
function by adding quasi-chemical and electrostatic non-native interactions. 
In both cases, the stochastic evolution is followed up
to the formation of about  35\% of the native contacts and the knotted
topology of the partially-folded structures is monitored.

We find that knot occurrence is negligible for both the AOTCase and
OTCase when the energy function  favoring only native contacts is
used. A dramatic difference in knotting propensity is instead seen when
the quasi-chemical non-native interactions are added. In this
case, the level of self-entanglement remains negligible for the OTCase but is greatly enhanced
for the AOTCase, which is natively knotted. Notably, all proper and improper knots formed in
these early folding stages have the correct (native) trefoil topology
and chirality. 

The analysis of the repeated knotting/unknotting events observed in
the simulated trajectories indicates that knotting usually results from
the threading of the C-terminal through loops present in the loose
protein globule. The threading events are  favored by the effective
(non-native) attraction of the C-terminus to other protein
regions. Additional folding simulations carried out for ``in silico
mutants'' of the AOTCases, provide further support for the effect of the
C-terminus chemical composition on the knotting propensities.

\section*{Results/Discussion}

The present investigation of knot formation in proteins that are only
partially folded is focused on two transcarbamylases: AOTCase, PDB
entry 2g68, and OTCase, PDB entry 1pvv.  The monomeric units of these two proteins have
nearly the same length (332 and 313 amino acids, respectively) and are
evolutionarily related. This is established from their significant
sequence homology (same CATH code, 3.40.50.1370 \cite{CATH} and
sequence identity equal to $\sim$40\%).  Despite their
statistically-significant structural alignability (Mistral p-value
$\approx 10^{-6}$ \cite{micheletti_plos,mistral}) they have a
different knotted topology \cite{virnau_plos,micheletti_plos}, see
Fig.~\ref{Figure_1}.  Specifically, the OTCase is unknotted, while the AOTCase
contains a right-handed trefoil knot. The knotted region, established using the 
algorithm of Ref.~\cite{knot_server}, corresponds to the K172-G255
segment and is at a distance of about 80 amino acids from the C terminus, which 
is the nearest one in sequence. Because the biological units of the two proteins have a different 
oligomeric state it is assumed that, consistently with other multimeric knotted proteins, the folding of the
monomeric units precedes their assembly. We shall therefore limit considerations to the
monomeric units.

\subsection*{The model}

For the purpose of our study, both proteins are described with a
simplified structural model, where each amino acid is represented by
one interaction center, coinciding with the C$_\alpha$ atom.

\begin{figure}[!ht]
\begin{center}
\includegraphics[width=12cm]{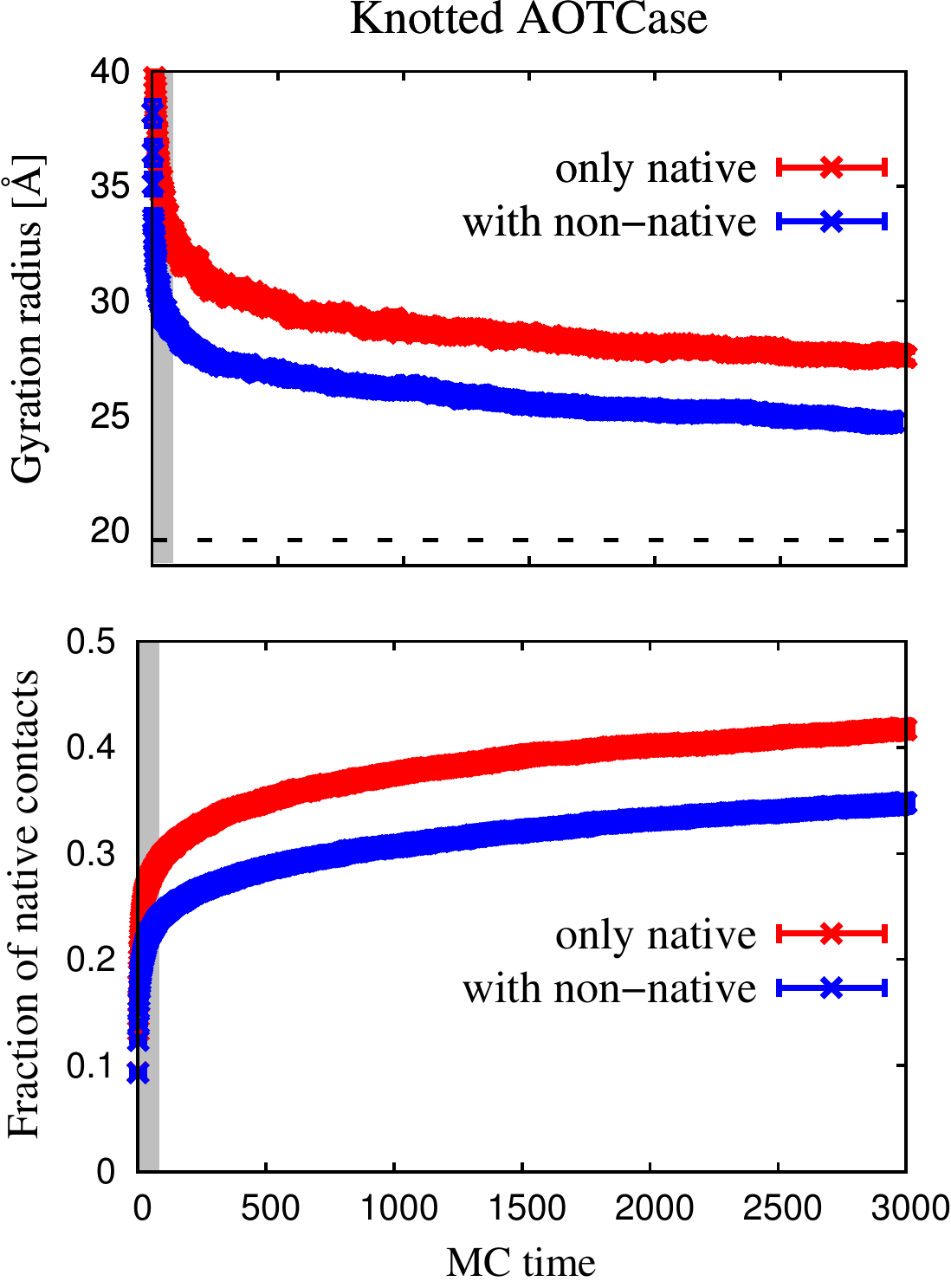}
\end{center}
\caption{{\bf Monte Carlo time evolution of the average gyration radius (top panel) and fraction of
    native contacts (bottom panel) for the knotted  AOTCase}.
The data obtained with the purely native-centric model are shown in red while those obtained with the added 
non-native interactions are shown in blue. In the top panel, the dashed line indicates gyration radius in the
  native state. Here, and in subsequent related figures, data points represent an average over the 150 trajectories and the associated statistical uncertainty is represented by the spread of the curves on the coordinate axis.  One unit of MC time corresponds to 100 attempted MC moves per amino acid. The gray region denotes the initial MC evolution where global pivot moves are employed to relax the initial fully-extended conformation.}
\label{Figure_2}
\end{figure}

The folding dynamics is simulated using the stochastic Monte Carlo (MC) approach 
introduced and validated in Ref.~\cite{fast_MC_algorithms}. The MC evolution  entails a brief relaxation from the
initial fully-extended state using both local and non-local moves. After this stage, indicated with a shaded region 
in Fig.\ref{Figure_2} and related ones, the MC dynamics proceeds exclusively through local moves. The moves amplitude is sufficiently small that no chain crossings occurs, as verified {\em a posteriori.
The stochastic scheme is analogous to the kink-jump dynamics~\cite{kink-jump1}  which, by virtue of the 
local character of the moves, is known to provide a physically-viable description of biopolymers' kinetics in thermal equilibrium~\cite{fast_MC_algorithms,kink-jump2}.}

The folding simulations are carried out using 
two alternative energy functions. The first one promotes exclusively the formation of native
contacts between pairs of amino acids. Following Refs. \cite{best_hummer,kim_hummer}, the strength of the attractive
interaction between any given native amino acid pair is derived from the
strength of their hydrogen bonding in the native state and is expressed in thermal units, $k_B\, T$, at the
nominal Monte Carlo temperature of $T=300$K. The second energy function includes electrostatic and 
non-native pairwise interactions in addition to the native-centric potential. 
The relative strength of the non-native interactions
is set according to the quasi-chemical potentials of Miyazawa and
Jernigan (MJ) \cite{miyazawa_jernigan} that reflect the statistical
propensity of amino acid pairs to be in contact in proteins'
native states. Following, again, Ref.s~\cite{best_hummer,kim_hummer}, the
average strength of the added non-native interactions is set to one tenth of the native one.
No change to the MC temperature was made using this second potential because previous 
studies have shown that the addition of non-native interactions modifies the
effective temperature of the system by less than $10\%$~\cite{PNASDRP}. Consistently with this 
fact we have verified that the native states of both AOTCase and OTCase remain stable when evolved
with the MC scheme using the second type of potential. In fact, the asymptotic value for average fraction of native contacts was $q\approx90$\%. The fact that this value is smaller than the asymptotic one,  $q\approx 95$\%, of the first type of potential
indicates that the strength non-native interactions while weak enough to maintain stable the native state, can nevertheless compete with the native attractive interactions. The addition of the quasi-chemical interactions are therefore expected to be capable of 
accounting for sequence-specific non-native contact propensities in the partly-folded state. The sequence-dependent character of the non-native interaction differentiates the present approach from the   early one of Wallin {\em et al.}~\cite{shakhnovich} where the knot-promoting effect of non-native interactions was probed by systematically introducing attractive interactions between various pairs of segments of YibK.

For each of the two proteins and for each of the two
energy functions, we generated $\sim$150 MC trajectories each
consisting of $\sim 3 \times 10^5$ MC moves 
 per amino acid (corresponding to 3000 units of MC time reported
in  Fig.~\ref{Figure_2} and  Fig.~\ref{Figure_3}), 
for a total of 75,000 CPU hours. The acceptance rate of the local MC moves for both proteins
was nearly constant throughout the simulation and approximately equal to 50\%.

The entanglement of the structures sampled during the MC evolution was established using
a combination of two unrelated knot detection schemes, which are described in the Materials and Methods. 
This choice was made to maximize the robustness of the criterion used to establish the knotted state of an open chain  - which is mathematically properly defined only after the chain termini are joined by a segment or arc thus giving a circular chain (chain closure operation).
As described in the Materials and Methods section we used many alternative closures for each chain and adopted a majority 
rule to single out chains that have non trivial self-entanglements. These chains can correspond to proper, fully developed knots (fully accommodated in the knotted structures) as well as improper ones (partly accommodated in the closing arcs).


\subsection*{Early folding kinetics}

\begin{figure}[!ht]
\begin{center}
\includegraphics[width=12cm]{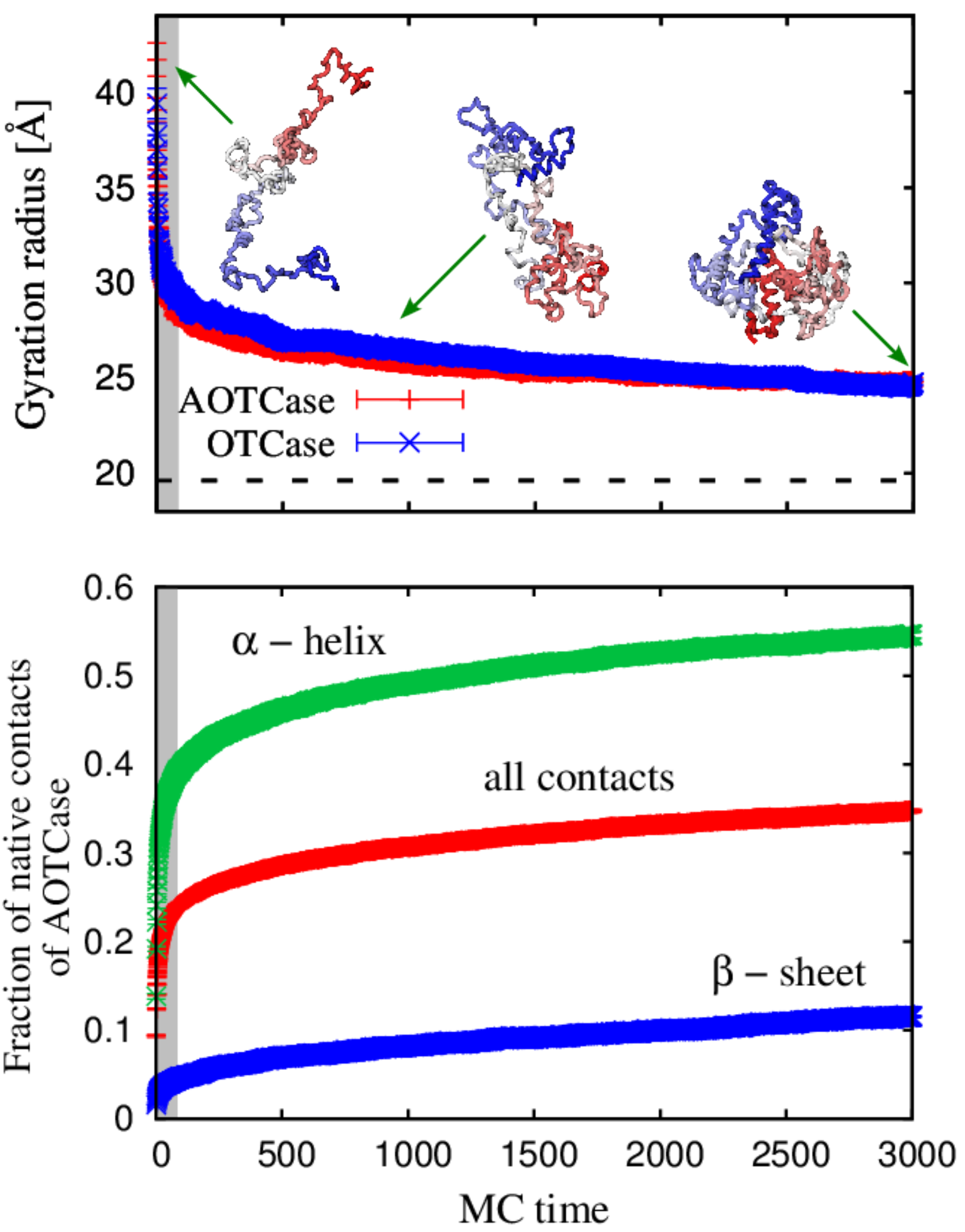}
\end{center}
\caption{{\bf Top panel: Monte Carlo time evolution of the average gyration radius
of the knotted  AOTCase (red) and the unknotted  OTCase (blue) in the model including both native and non-native interactions.}
The dashed line indicates gyration radius in the native state. Each data point is an average over the 150 trajectories and its statistical uncertainty is represented by the spread of the curves on the coordinate axis.
The overlaid structures are instantaneous C$_\alpha$ traces of AOTCase; the N and C termini are colored in red and blue, respectively.
 Bottom panel: Monte Carlo evolution of the fraction of native contacts of AOTCase. The overall fraction is shown in red, while the 
the fraction of formed native contacts involved in $\alpha$-helices and $\beta$-sheets are shown in green and blue, respectively.}
\label{Figure_3}
\end{figure}

In order to illustrate how the folding process differs for the two potential energy functions, we show in Fig.~\ref{Figure_2} 
the evolution of the gyration radius (upper panel) and the 
fraction of native contacts (lower panel) of the 
knotted AOTCase, obtained by employing these two potentials.  It is seen that the effect of the 
non-native and electrostatic interactions results in the higher compactness 
of the protein globule, since the gyration radius in this case lies systematically 
below the one calculated with the purely native-centric model.
At any given stage of the MC evolution, the fraction of established native contacts is systematically
lower for non-native interactions thus indicating that the latter
introduce frustration that slows down the folding dynamics. Analogous results hold for the unknotted OTCase, see top panel of Fig.~\ref{Figure_3}.

The data in Fig.~\ref{Figure_3} represent how the average radius of
gyration and fraction of formed native contacts evolve in the course
of the folding simulations of the two transcarbamylases. It is seen that, over the
duration of the simulation, the average fraction of formed native
contacts grows to $q \approx 35$\% on average (the maximum value in all trajectories was $q \approx 50$\%. Overall, the probability of formation of
native contacts decreases with the sequence separation of the amino acid 
pair, so that native $\alpha$-helical contacts are substantially
more probable than inter-strand ones \cite{finkelstein1}.

From the steady increase of the fraction of native contacts, $q$, it is
extrapolated that folding completion would occur on time-scales at
least ten times larger than considered here. 
From the same figure it emerges that the radius of gyration gradually decreases to about 25 $\text{\AA}$ 
which is 30\% larger than the native one, consistently with the less
compact character of the partially folded states.

To characterize the overall propensity of the two proteins to form
knots while still largely unfolded, we extracted configurations at
regular intervals of the simulations and analyzed their topological
state.

\subsection*{Non-native interactions and self-entanglement}

\begin{figure}[!ht]
\begin{center}
\includegraphics[width=12cm]{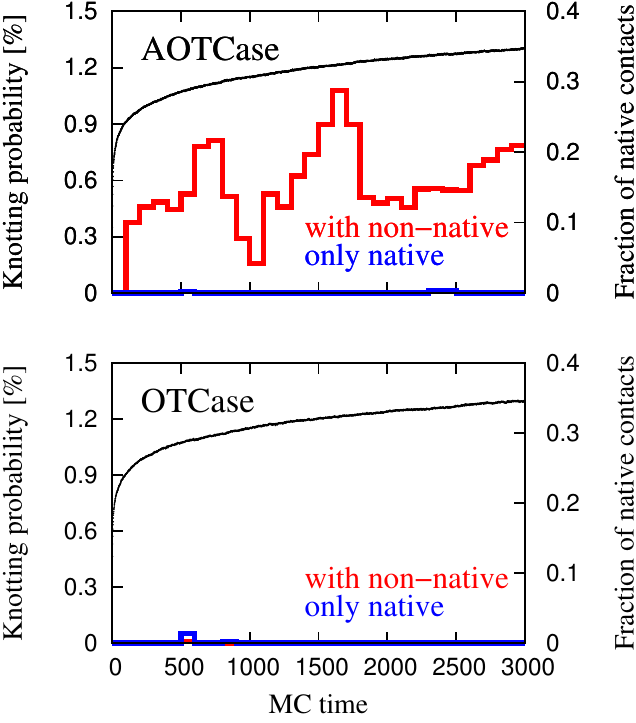}
\end{center}
\caption{{\bf Monte Carlo time evolution of the average knotting probabilities and fraction of native contacts of the natively-knotted
 AOTCase (top panel) and of the unknotted  OTCase (bottom panel).} The thin black curve shows the average fraction of formed native contacts. The knotting probabilities observed for the purely native potential and for the added non-native interactions are shown with thick blue and red lines, respectively.}
\label{Figure_4}
\end{figure}

The knotting propensity of AOTCase, which is natively knotted, is illustrated
in the upper panel of Fig.~\ref{Figure_4} which portrays the fraction of 
configurations that are properly or improperly knotted during a given time-window of the MC 
dynamics.

It is seen that when the purely native-centric potential is used, the
knotting propensity is always negligible. However, when
non-native interactions are added, the number of knotted
configurations increases to a definite fraction of the total: at the end of the simulations, when only 35\% of the native
contacts are formed, the fraction of conformations that are 
properly or improperly knotted is about 1\%. Notice that this value is comparable with the yield of purely 
native-centric coarse-grained models where pathways with a late-stage formation of knots are typically
observed \cite{onuchic1}. Within the limitations of coarse-grained approaches, the results
indicate that sequence-dependent non-native interactions can produce a detectable fraction of configurations with the correct native entanglement already at early folding stages.

The analysis was repeated for the natively-unknotted OTCase and
the results are shown in the lower panel of Fig.~\ref{Figure_4}. The
contrast with the case of the ATOCase is striking. In fact, the
knotting probability is negligible not only for the purely-native
case, but remains so even when non-native interactions are introduced.

The results indicate that the sequence-dependent non-native
interactions -- promoted by the quasi-chemical potential -- increase
dramatically the incidence of knots in the partially folded state of
the AOTCase compared to the OTCase. Equivalently, the model calculations indicate
that the two related natively-knotted and unknotted proteins have
different knotting propensities already in the partially folded state,
and that this difference can be ascribed to sequence-specific
non-native interactions.

\subsection*{ AOTCase knotting events}

To further elucidate the mechanisms responsible for these differences
we monitored several parameters in the course of the simulations of the two 
transcarbamylases. More specifically, we searched for systematic differences
in the interactions that the termini of these proteins establish with other parts of the
peptide chain and for preferential locations of fully-developed knots.

\begin{figure}[!ht]
\begin{center}
\includegraphics[width=12cm]{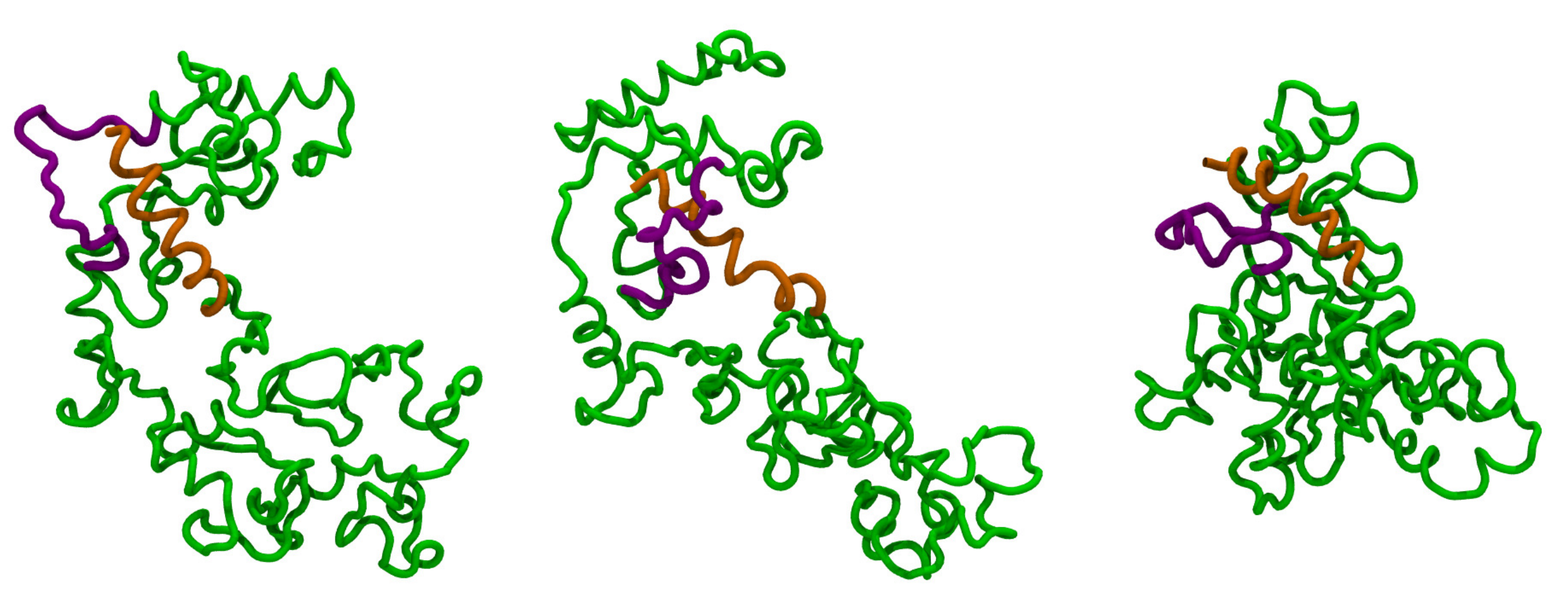}
\end{center}
\caption{{\bf Three configurations of the knotted AOTCase
in the presence of non-native interactions that illustrate a knotting-unknotting 
event.}
Three MC generated coarse-grained configurations (subsequent in MC time):
unknotted (left and right) and knotted (middle). The knot results from the
threading of a loop (in cyan) by part of the  C-terminal $\alpha$-helix (colored in orange).}
\label{Figure_5}
\end{figure}

In this respect, it is important to point out that, in the course of
the simulations for the AOTCase with non-native interactions, the process of
knot formation is not irreversible. As illustrated in
Fig.~\ref{Figure_5}, trefoil knots can be formed and untied
during each simulation. In particular,  fully-developed knots can persists for up to one tenth of
the duration of our simulations. Notably, only knots with the correct (i.e. native) chirality are observed.  
 
 By analysing the configurations preceding and following the knotting/unknotting events it is found
that the change in topological state is typically caused by the
threading of the helical C-terminus through loops formed by various
protein regions. 

Among the independent knotting events we identified six for which the stochastic closure returned
a non-trivial topology in at least 80\% of the cases or the portion accommodating the knot had a depth of at least 20 amino acids.
The analysis of the loop threading events leading to such persistent fully-developed knots
showed that they involved the C terminus interacting with amino acids 55-65,
 90-110, or  125-155. Only in one case it was observed the threading of the N-terminal through a loop involving segments 252-277.
 
\begin{figure}[!ht]
\begin{center}
\includegraphics[width=12cm]{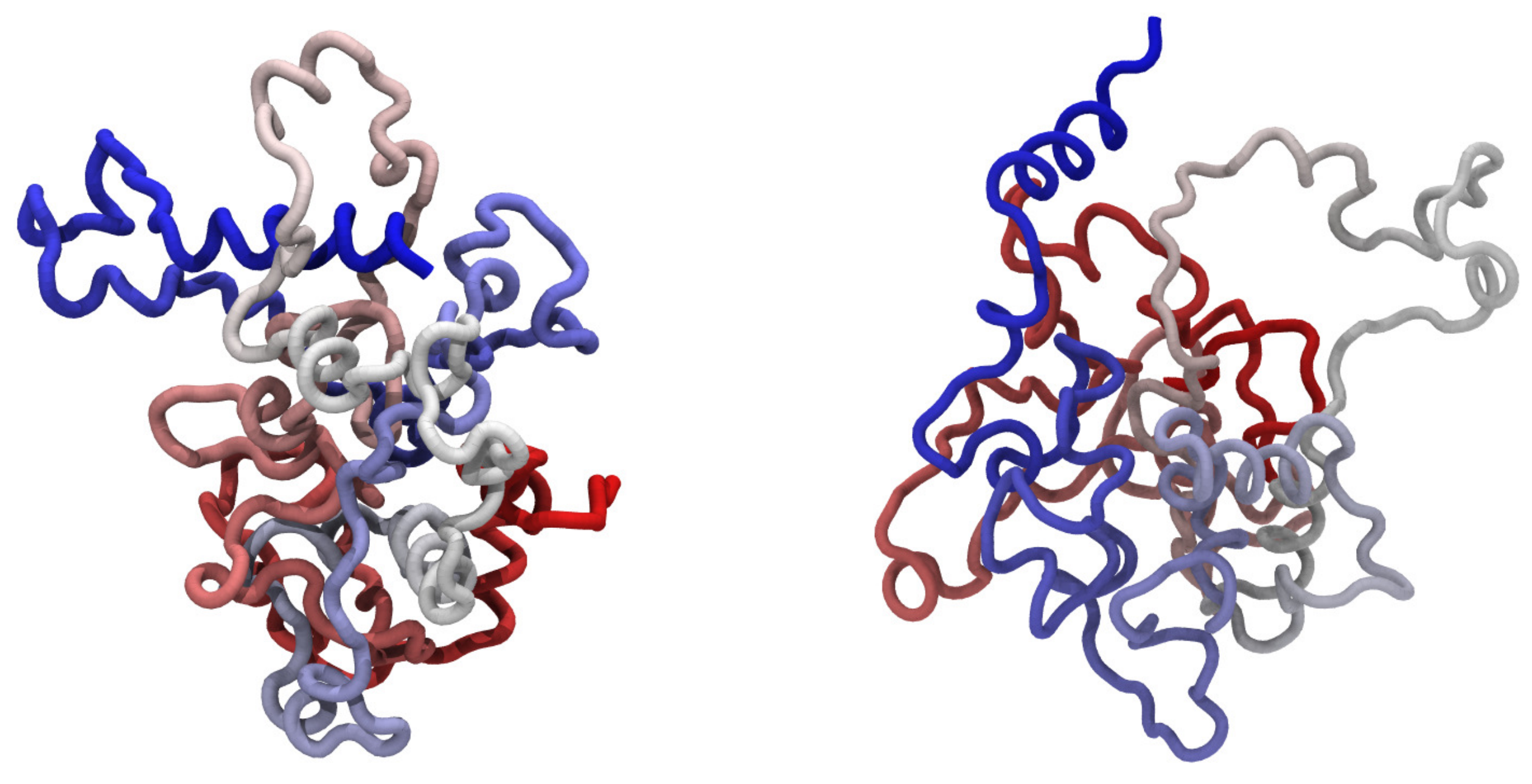}
\end{center}
\caption{{\bf Typical protein configurations obtained in the presence of non-native interactions: that of the knotted 
AOTCase (left panel) and of the unknotted OTCase
(right panel).} The structures illustrate the tendency of the C-terminal $\alpha$-helix (colored in blue)
to point towards the protein  globule for the knotted protein and away from it for the 
unknotted one.}
\label{Figure_6}
\end{figure}

Representative configurations are shown in the left panel of
Fig.~\ref{Figure_6} where it is clear that the helical C-terminus of the AOTCase typically points
towards the rest of the protein, and the stiffness of the helix
facilitates the threading of various loop regions.

At variance with the above situation, for both potential energy
functions the OTCase C-terminus is typically exposed to the solvent
and pointing away from the rest of the protein, as in the example shown
in the right panel of Fig.~\ref{Figure_6}. The terminus also shows a lower 
propensity to form $\alpha$-helices. An analogous situation is found for the AOTCase 
in the presence of only native interactions.

\begin{figure}[!ht]
\begin{center}
\includegraphics[width=12cm]{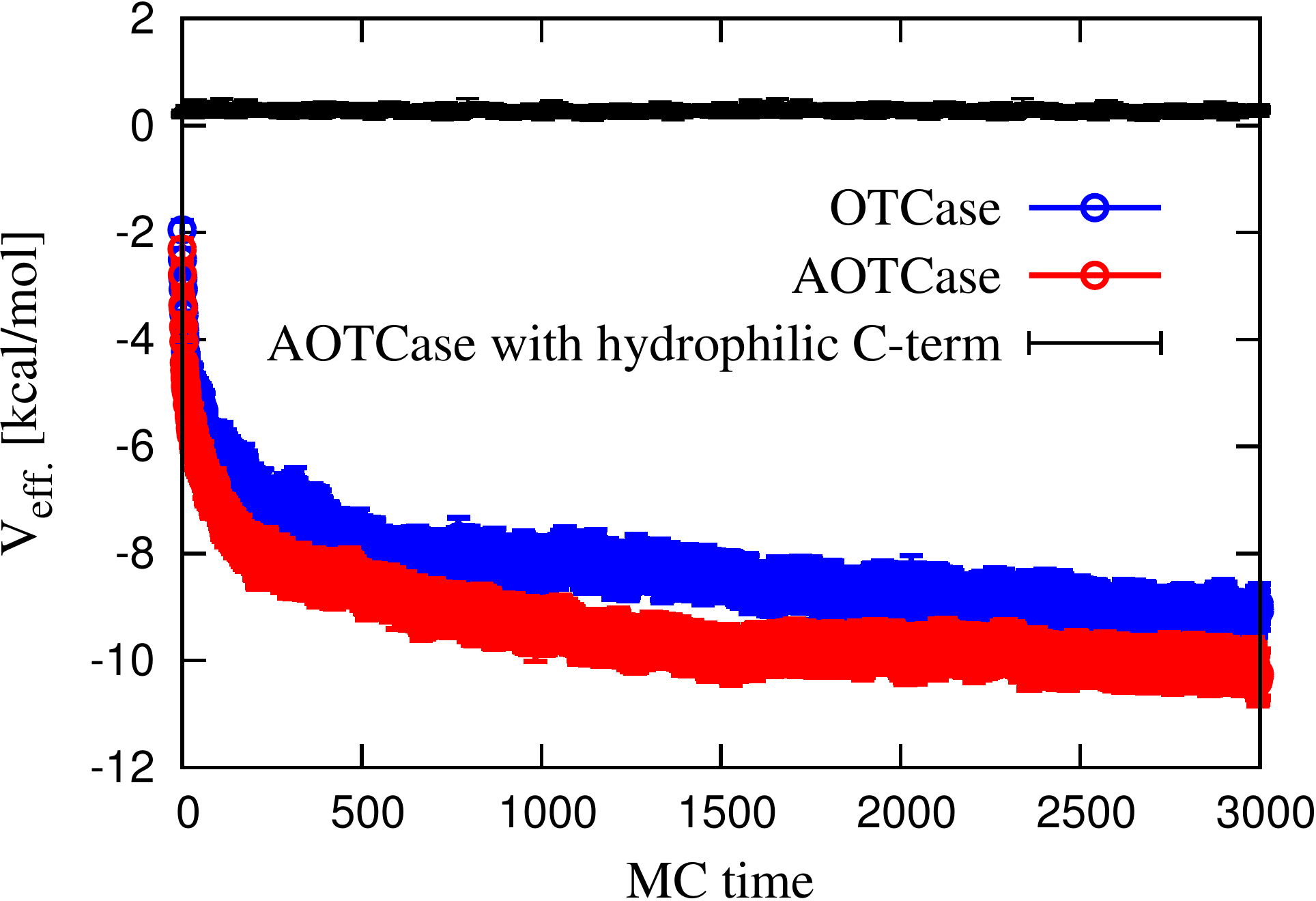}
\end{center}
\caption{{\bf Monte Carlo time evolution of the average quasi-chemical interaction energy between 
the $\alpha$-helix and the rest of the chain, in the knotted  AOTCase 
and unknotted  OTCase.} 
The upper curve shows the same quantity for a mutant of the natively knotted protein in 
which all residues in the C-terminal $\alpha$-helix are replaced by hydrophilic residues.}
\label{Figure_7}
\end{figure}

The effect is quantitatively illustrated in Fig.~\ref{Figure_7},
which portrays the average strength of the non-native interaction
potential energy between the amino acids in the C-terminal
$\alpha$-helix of the AOTCase or the OTCase and all the other residues in the
chain.  It is seen that the average non-native attraction is
consistently stronger in the case of the knotted protein by about
15\%, which is a sizable amount considering that the average is taken
over all sampled structures, irrespective of their compactness and
knotted topology.

The same figure also illustrates that, if the amino acids in the
C-terminal $\alpha$-helix of protein AOTCase are replaced by neutral
hydrophilic residues (GLU, GLN, ASN), then the average non-native
interaction energy becomes slightly repulsive, explaining the tendency of 
the $\alpha$-helix to point outwards. This provides an additional argument 
in support of the picture where the hydrophobicity character of the residues 
plays a role in the folding of knotted proteins~\cite{shakhnovich}.

\begin{figure}[!ht]
\begin{center}
\includegraphics[width=12cm]{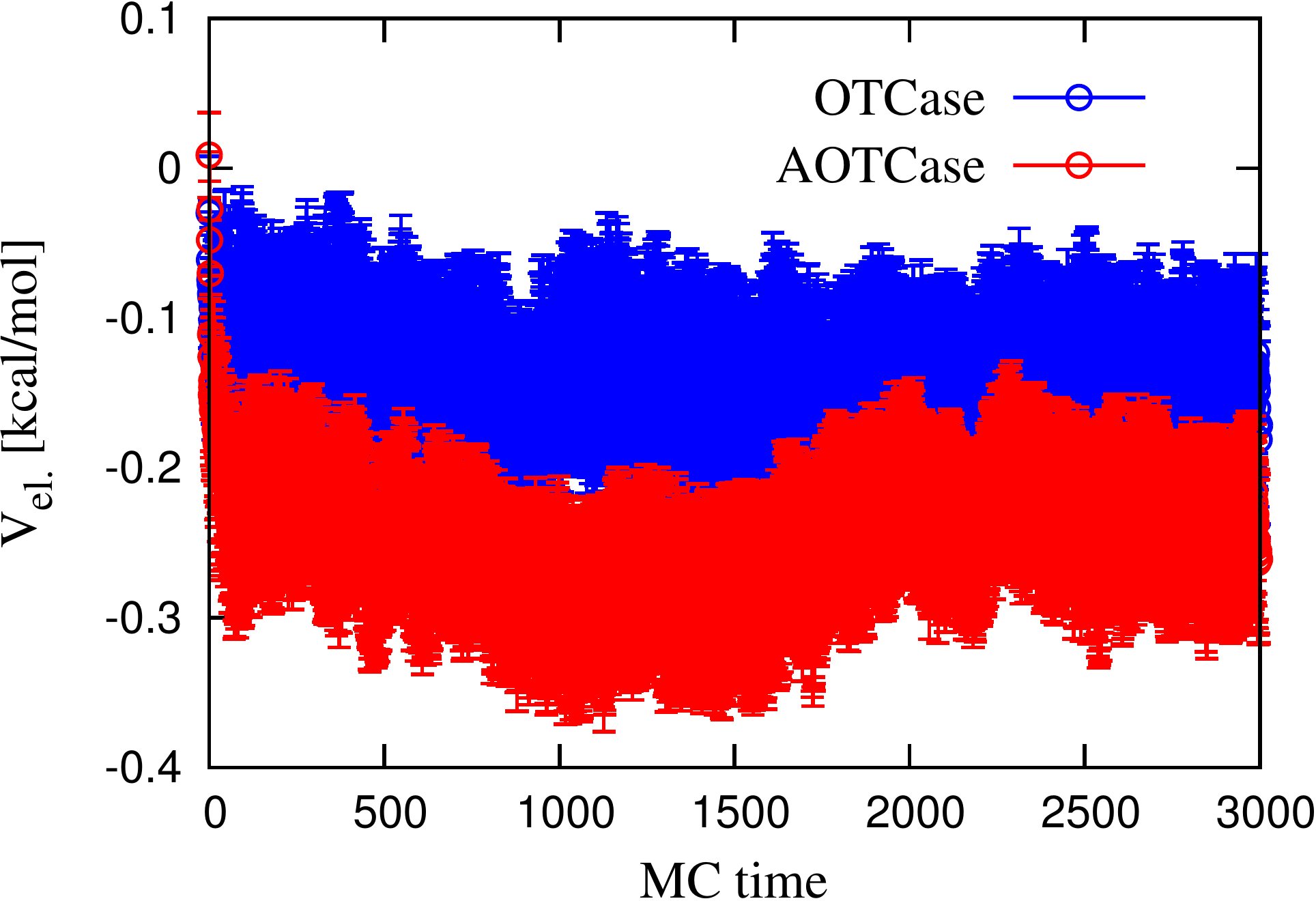}
\end{center}
\caption{{\bf Monte Carlo time evolution of the average electrostatic interaction energy between 
the $\alpha$-helix and the rest of the chain, in the knotted  AOTCase
and unknotted  OTCase.}}
\label{Figure_8}
\end{figure}

Finally, in Fig.~\ref{Figure_8} we plot the average electrostatic potential
energy between the $\alpha$-helix and the rest of the chain and show that, 
overall, the  electrostatic interaction is about two orders of magnitude 
smaller than that of the quasi-chemical potential. 

\begin{figure}[!ht]
\begin{center}
\includegraphics[width=12cm]{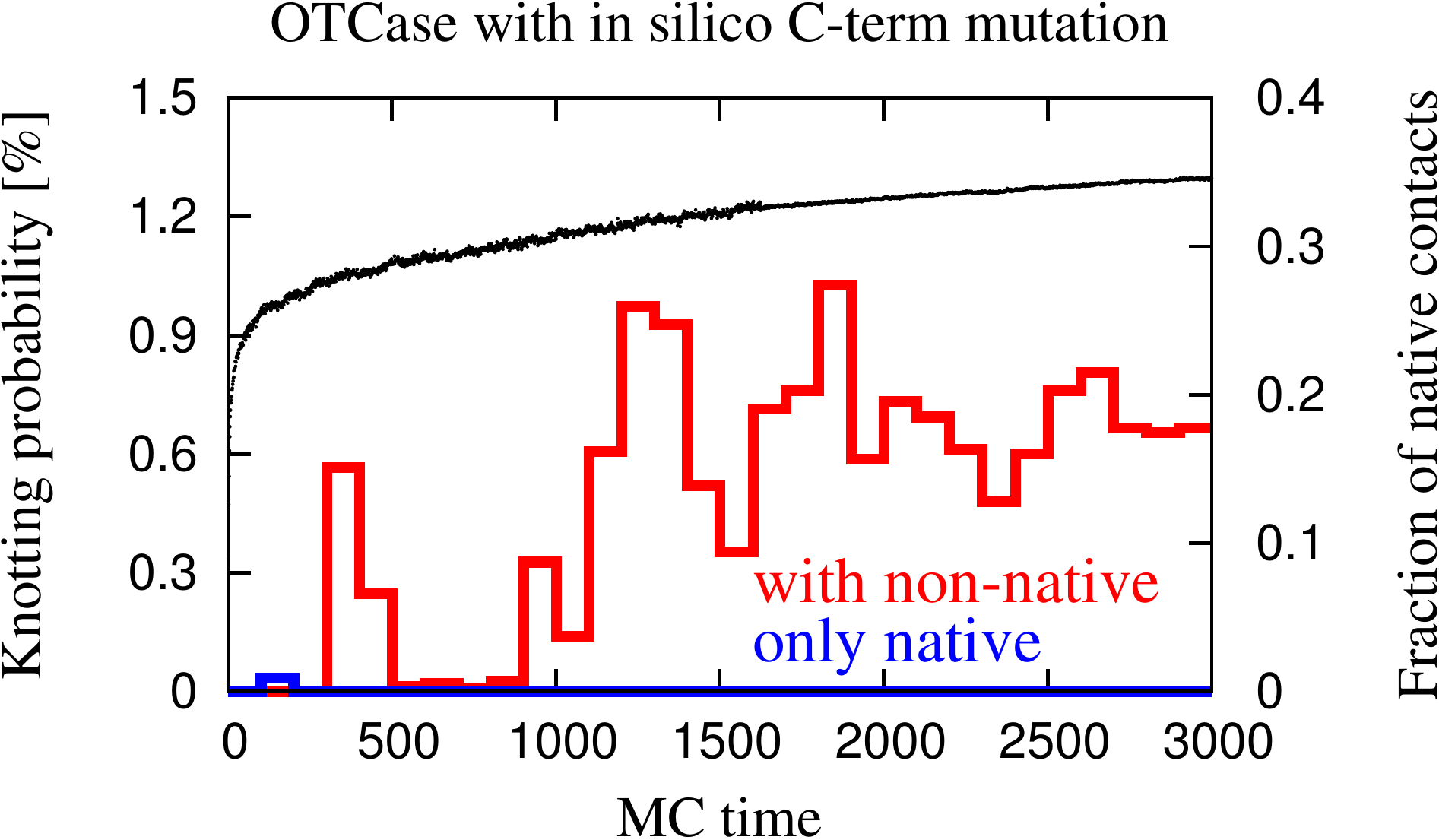}
\end{center}
\caption{{\bf Monte Carlo time evolution of the average knotting probabilities and fraction of native contacts of the "in silico" mutant of the natively-unknotted  OTCase}. The knotting probabilities observed with  the 
purely native-centric model and with the added non-native interactions are shown in blue and red, respectively.
The black curve shows the fraction of native contacts in the presence of non-native interactions.}
\label{Figure_9}
\end{figure}

To further clarify the role of the C-terminal $\alpha$-helix in the knotting of
the partially-folded states of the AOTCase, we have carried out another set
of simulations on a mutant protein. The last 25 residues of the OTCase (which are involved
in the C-terminal $\alpha$-helix) were substituted with the same number of
residues that form the C-terminal $\alpha$-helix in the knotted
AOTCase. Apart from such a substitution, all other attributes, namely the length of the chain and
the native contact map, were identical to the unknotted protein.  
The outcome of the simulation is summarized in Fig.~\ref{Figure_9}, from
which we can conclude that the mutant protein forms a substantial
fraction of knotted configurations in the presence of non-native
interactions in its early stage of the folding.

\subsection*{Summary and conclusions}

A simplified coarse-grained protein model was used for a comparative
study of early folding stages of two evolutionarily related
transcarbamylase proteins: an AOTCase (PDB id 2g68) and an OTCase (PDB
id 1pvv). The two proteins are well-alignable in sequence and
structure and yet possess differently knotted native states: AOTCase is trefoil-knotted while OTCase is unknotted.

The role of sequence-dependent non-native interactions in promoting the correct native
topology in the early folding stages was investigated by using two different
energy functions in the simulations: one with a purely native-centric
potential energy ( favoring only native contact interaction) and one
with the additional contribution of quasi-chemical and electrostatic non-native
interactions.

We found that in the absence of quasi-chemical interactions, neither
protein shows an appreciable propensity to self-entangle in the early
folding stages.  However, once non-native interactions are
introduced, the natively-knotted AOTCase does show a strongly enhanced
propensity to form proper and improper knots with the correct native topology and chirality.
By contrast, the knot enhancement effect
is completely absent in the natively-unknotted OTCase.

Inspection of the ensemble of folding trajectories of the two proteins
suggest that knotting in the partly unfolded AOTCase mostly
results from the approach of the hydrophobic C-terminal $\alpha$-helix
to other regions of the protein which are eventually threaded
through. Because of the non-compact character of the partly-unfolded
structures, the $\alpha$-helix in the C-terminal can also retract from
the threaded regions, so that various events of formation/disruption of proper and improper knots can be
observed in a given trajectory.

\noindent 

The influence of the C-terminal region on the different knotting
propensity of the two carbamylases is further supported by the folding
simulations for an ``in silico mutant'' of the OTCase obtained by
replacing the C-terminal sequence with the one of the knotted
AOTCase. In fact, the folding simulations based on the quasi-chemical
non-native interactions yielded a portion of knotted structures
similar to the natively-knotted AOTCase, from which the C terminus was
taken.

We recall that, the seminal study of Wallin {\em et al.} \cite{shakhnovich}, which was based on a coarse-grained model of YibK with {\em ad hoc} non-native interactions, had suggested the relevance of non-native interactions for steering the early formation of the native knotted topology (by the threading of a loop by the C-terminal). Our results, which are consistent with these conclusions, provide additional elements to the picture by taking advantage of a coarse-grained framework where the folding of two carbamylases with similar structures, and yet different knotted topology are compared on equal footing. In particular our results indicate that non-native interaction propensities that are encoded in the primary sequence, can favor or disfavor the formation of knots already at early folding stages.

The necessarily limited  scope of the coarse-grained approach used here does not clarify whether additional pathways leading to a late-stage formation of knots can be  present in transcarbamylases. It would be most interesting to address this standing issue in future studies within the present, or alternative comparative schemes.

\section*{Materials and Methods}

\subsection*{Coarse-Grained Model}

We have adopted the coarse-grained model developed in 
Ref.s~\cite{best_hummer,kim_hummer}, in which the effective degrees of 
freedom are the amino acid residues, represented by a spherical bead 
located at the position of the corresponding C$_{\alpha}$ atom. The 
potential energy of the model consists of bonded and non-bonded terms. 
The bonded part of the of the potential consists of stretching 
potentials of the pseudo-bond \mbox{C$_{\alpha}$-C$_{\alpha}$} and pseudo-angle 
\mbox{C$_{\alpha}$-C$_{\alpha}$-C$_{\alpha}$}, as well as potential for 
pseudo-torsions \mbox{C$_{\alpha}$-C$_{\alpha}$-C$_{\alpha}$-C$_{\alpha}$}:
\begin{equation}
V_{\text {bonded}} = V_{\text {bond}} (r_{ij})
                   + V_{\text {angle}} (\theta_{ijk})
                   + V_{\text {torsion}} (\varphi_{ijkl})\,.
\end{equation}
%

\vspace*{2mm}

\indent $\bullet$ The bond-stretching potential has the form
\begin{equation}
V_{\text {bond}} (r_{ij})=\frac{1}{2}k(r_{ij}-r_0)^2\, ,
\label{equation_bond}
\end{equation}
where $r_{ij}$ is the distance between the residues $i$ and $j$,
while $r_0$ is the equilibrium length of the C$_{\alpha}$-C$_{\alpha}$ 
pseudo-bond.
%
\newline
\vspace*{2mm}

\indent $\bullet$ The double-well pseudo-angle potential is given by

\begin{equation}
V_{\text {angle}}(\theta_{ijk}) =  -\frac{1}{\gamma} { \ln} \Big[
e^{-\gamma (k_{\alpha}(\theta_{ijk}-\theta_{\alpha})^2+\epsilon_{\alpha})} 
{ +} \,e^{-\gamma k_{\beta}(\theta_{ijk}-\theta_{\beta})^2} \Big],   
\label{equation_angle} 
\end{equation}
where $\theta_{ijk}$ is the pseudo-angle formed by the residues $i$, $j$ 
and $k$, while $\theta_{\alpha}=92^{\circ}$ and 
$\theta_{\beta}=130^{\circ}$ are the equilibrium values of the helical 
and the extended pseudo-angles, respectively. 
%
\newline
\vspace*{2mm}

\indent $\bullet$ The torsion-angle potential for pseudo-torsions is of the form
\begin{equation}
V_{\text {torsion}}(\varphi_{ijkl}) = 
\sum_{\text {n}=1}^{4}
\Big[
1+\cos(\text{n}\varphi-\delta_{\text n})
\Big] 
V_\text{n},
\label{equation_torsion}
\end{equation}
with $\varphi_{ijkl}$ being dihedral angle between the planes
($i$,$j$,$k$) and ($j$,$k$,$l$). The constants $\delta_{\text{n}}$ 
and $V_{\text n}$ are depending only on the type of the middle two 
residues $j$ and $k$ and are adopted from Karanicolas and 
Brooks~\cite{karanicolas_brooks}.
%

All the values of the spring and energy constants appearing in the
expressions of the bonded energy terms in Eqs.~(\ref{equation_bond}), 
(\ref{equation_angle}) and (\ref{equation_torsion}) can be found in 
Ref.~\cite{kim_hummer}.

The non-bonded part of the potential consists of three terms, comprising 
native, non-native and electrostatic interactions, respectively: 
%
\begin{equation}
V_{\text{non-bonded}} = \sum_{\text{native\,} (i,j)} V_{\text {Go}}(r_{ij}) 
                       + \sum_{\text{non-native\,} (i,j)} V_{\text{eff.}}(r_{ij}) 
                       + \sum_{\text{all\,} (i,j)} V_{\text{el.}}(r_{ij})\,.
\label{equation_general} 
\end{equation}

\vspace*{2mm}

\indent $\bullet$ $V_{\text{Go}}(r_{ij})$ denotes the G$\overline{\textrm{o}}$-type potential developed by   
Karanicolas and Brooks~\cite{karanicolas_brooks}. Within such an approach, 
the native contact map is defined on the basis of the network of the 
hydrogen bonds in the native state, as well as on the degree of 
the proximity of the backbone atoms side-chains. Namely, two residues 
are defined to be in the native contact if the hydrogen bond between 
them is stronger than -0.5 kcal/mol, or if any of their non-hydrogen side-chain 
atoms are within the distance of 4.5 $\text{\AA}$, in the native state. We do not consider contacts between residues with a distance in sequence smaller than 3 amino-acids. 
These interactions are described by the following functional form:
\begin{equation}
V_{\text{Go}}(r_{ij}) = \varepsilon_{ij} 
\bigg[ 13 \Big(\frac{\sigma_{ij}}{r_{ij}}\Big)^{12}
-18\Big(\frac{\sigma_{ij}}{r_{ij}}\Big)^{10}
+4\Big(\frac{\sigma_{ij}}{r_{ij}}\Big)^{6}
\bigg],
\end{equation}
where $\sigma_{ij}$ is the native-state separation of residues $i$ and $j$. The 
strength of the interaction $\varepsilon_{ij}$ is chosen to be that of their 
hydrogen bond, for the residues that are hydrogen-bonded in the native state, while 
for the side-chain interacting residues it is a value proportional to
the corresponding MJ contact 
potential~\cite{miyazawa_jernigan} and it was suitably renormalized 
in order to match the hydrogen bond native contact energy scale. 
The residues in $\beta$-sheets or hairpins are often in contact via
multiple hydrogen bonds, so in order to stabilize these structures within 
the G$\overline{\textrm{o}}$-model, the additional network of weaker hydrogen bonds (having the strength 
$\varepsilon_{ij}/4$) involving the neighboring pairs $(i-1,j)$, 
$(i,j+1)$, $(i+1,j)$ and $(i,j-1)$ was introduced around each residue pair $(i,j)$ 
that was found to interact via either two hydrogen bonds or a hydrogen bond and 
a side-chain contact. 
\newline
\vspace*{2mm}

\indent $\bullet$ Depending on the amino acid type of the residues $i$ and $j$, the non-native
interactions in the model developed by Kim and Hummer in Ref.~\cite{kim_hummer}
can be both attractive and repulsive. Repulsive interactions are 
applied between amino acid pairs that interact less favorably with each other than
with the solvent and {\it vice versa}. For a pair of residues that experiences
an effective attractive interaction with a strength $\varepsilon_{ij} < 0$, the non-native 
interaction potential is given by
\begin{equation}
V_{\text{eff.}}(r_{ij}) = 4|\varepsilon_{ij}| 
\bigg[ 
\Big(\frac{\sigma_{ij}}{r_{ij}}\Big)^{12}-
\Big(\frac{\sigma_{ij}}{r_{ij}}\Big)^{6}
\bigg]\,,
\end{equation}
with $\sigma_{ij} = 1/2 (\sigma_i+\sigma_j)$, where $\sigma_i$ and $\sigma_j$
are van der Waals radii of the residues $i$ and $j$. 

For pairs of residues that effectively repel each other, so that $\varepsilon_{ij} > 0$, 
the non-native potential energy function takes the following form:
\begin{equation}
V_{\text{eff.}}(r_{ij}) = \left\{\begin{array}{rl}
                                    4 \varepsilon_{ij} \bigg[
                                      \Big(\frac{\sigma_{ij}}{r_{ij}}\Big)^{12}-
                                      \Big(\frac{\sigma_{ij}}{r_{ij}}\Big)^{6}
                                    \bigg] + 2 \varepsilon_{ij}, & r_{ij} < \sqrt[6]2\,\sigma_{ij}\\
                                   -4 \varepsilon_{ij} \bigg[
                                      \Big(\frac{\sigma_{ij}}{r_{ij}}\Big)^{12}-
                                      \Big(\frac{\sigma_{ij}}{r_{ij}}\Big)^{6}
                                    \bigg], & r_{ij} \geq \sqrt[6]2\,\sigma_{ij}
                                 \end{array}
                                 \label{veff}
                                 \right .
\end{equation}

The effective Lennard-Jones interaction strength $\varepsilon_{ij}$ between  
residues $i$ and $j$ in the coarse-grained model of Kim and Hummer~\cite{kim_hummer}
is defined as
\begin{equation}
\varepsilon_{ij} = \lambda (e_{ij} - e_0).
\end{equation}
The coefficients $e_{ij} $ are negative  and coincide with the entries of the MJ  matrix \cite{miyazawa_jernigan}, while $e_0$ is an offset parameter. Hence, Eq.~(\ref{veff}) defines a statistical knowledge-based 
potential which measures the preference of residue-residue interactions relative to residue-solvent 
interactions. The parameter $\lambda$ scales the strength of the Lennard-Jones 
interaction compared to the physical electrostatic interactions.
The two free parameters $\lambda$ and $e_0$ are fitted in order to 
correctly reproduce the binding affinity of the broad set of 
experimentally well-characterized protein complexes~\cite{kim_hummer}.
As an illustration, the effective interaction strength between the
neutral hydrophilic residues GLU and GLN is $0.08$ kcal/mol,
so that effective interaction is repulsive, while for the hydrophobic
residues ILE and LEU is $-0.46$ kcal/mol, so that
the effective interaction is attractive.

\vspace*{2mm}

\indent $\bullet$ $V_{\text{el.}}(r_{ij})$ is the long-range electrostatic interaction between
residues $i$ and $j$ and it is modeled by Debye-H$\ddot{\textrm{u}}$ckel-type of the potential 
\begin{equation}
V_{\text{el.}}(r_{ij}) =\frac{ q_i q_j}{4\pi \varepsilon_0 D}  ~\frac{\exp\left[-\frac{r_{ij}}{\xi}\right] }{ r_{ij}}\,,
\end{equation}
where $q_i$ and $q_j$ are the electrostatic charges of residues $i$ and $j$,
$\xi$ is the Debye screening length, $\varepsilon_0$ is dielectric vacuum
constant and $D$ is the relative dielectric constant of water in 
near-ambient conditions.
 

\subsection*{Numerical Simulation Details}

The non-native interactions in the force field of our coarse-grained 
model introduce frustration and make the potential energy surface quite 
rugged. As a consequence, even within such a simple model, molecular 
dynamics (MD) simulations of the folding for chains of several hundreds 
amino acids are very expensive. Within the simulation time intervals 
which were accessible to our computer resources, such MD trajectories 
did not allow to monitor significant changes in the chain conformations. 

To cope with this problem, we have simulated MC dynamics,
using an algorithm which combines different types of moves, namely:
\newline
\vspace*{2mm}

$\bullet$ local crankshaft moves \cite{binder_cranckshaft}, that consist of the rotation of 
a randomly selected single bead around the axis defined by its nearest 
neighbors. The angle of the rotation was randomly 
selected in the interval $\Delta \varphi_{max} = \pm 30^{\circ}$,\\

\vspace*{2mm}

$\bullet$ local end-point moves, in which the last 10 residues on both terminals are rotated rigidly with respect to the rest of the chain by  up to $30^{\circ}$ around a random axis passing through the most interior bead of the end segment.
\vspace*{2mm}

$\bullet$ local Cartesian moves, that involve the displacement of the coordinates of a single randomly 
selected bead in the chain, within a sphere of radius $0.15 \text{\AA}$,\\ 

\vspace*{2mm}

$\bullet$ global pivot moves \cite{pivot}, where one amino acid is picked at random and
the chain portion involving all amino acids with smaller (or alternatively larger) sequence index are rotated 
by up to $30^{\circ}$ around a random axis passing through the picked amino acid.

\vspace*{2mm}

The moves were accepted or rejected according to the standard Metropolis criterion.

MC algorithms based on local crankshaft and end moves are commonly
employed in the polymer physics~\cite{fast_MC_algorithms} to study 
\emph{dynamic properties}, since it is was shown that they can mimic the intrinsic dynamics 
of a polymer in solution~\cite{binder_cranckshaft} at a much lower computational 
cost of MD simulations~\cite{mc_for_comformational_sampling}. However, we emphasize
that since our computational scheme is based on the MC evolution, we can not establish 
a quantitative mapping between the simulation time and physical time. 

We have generated 150 independent MC trajectories for each system, 
starting from the same stretched coil configuration.  
In the early stage of the folding, these types of  moves were attempted 
with the constant probabilities of 0.8, 0.1 and 0.1, respectively. Such an 
algorithm generates a rapid collapse of the chain from a fully stretched 
configuration to one in which the gyration radius is reduced from about
70 $\text{\AA}$ to about 30 $\text{\AA}$ in about $10^{4}$ MC
steps per particle, that corresponds to 100 units of MC time
(see Fig.~\ref{Figure_3}). 
During such collapse, no knot was observed.
At the end of this stage, the acceptance rate of the global pivot moves typically 
dropped below 1\%. After this happened, we switched off the pivot moves, except from 
those involving few residues near the the two terminal of the chain. From this point 
on, the conformational changes of the chain are driven by the local crank-shaft and 
cartesian moves with an overall acceptance of about 50\%.

When computing the fraction native contacts $q$ along the calculated MC trajectories, we adopt a criterion according to which 
two residues with index difference $\ge 3$ are said to be in contact if the distance  of their $C_\alpha$ is less than $7.5$\AA.

\subsection*{Knot detection schemes}

The conformations visited during the Monte Carlo dynamics were topologically classified by computing the Alexander
determinants after suitable closure into a ring \cite{poly8}.

For robustness, two alternative closure schemes are used: the minimally-interfering closure~\cite{Min_entang_closure1} and a modified version of the stochastic one~\cite{stochasticClosure}. The minimally-interfering closure is used first because it is very computationally effective, in that it entails a single, optimally chosen closure. In case of positive knot detection we further validate the non-trivial entanglement  by performing 100 closures where each terminus is prolonged far out of the protein along a stochastically chosen direction, and the end of the prolonged segments are closed by an arc (that does not intersect the protein). The stochastic exit directions are picked uniformly among those that are not back-turning. Specifically, they must form an angle of more than 90$^\circ$ with the oriented segment going from each terminus to the $C_\alpha$ at a sequence distance of 10.  If the majority of the stochastic closures return non-trivial Alexander determinants, than the conformation is non-trivially entangled. Such conformations can correspond to both proper, fully developed knots,
and improper ones. The two can be distinguished using knot localization criteria~\cite{Min_entang_closure1,Min_entang_closure2}: proper knots are entirely accommodated within the original protein chain, while improper ones span the exit segments.
The knot type was determined using the scheme of Ref.\cite{poly6}, which is based on the KNOTFIND algorithm.

\section*{Acknowledgments}
We thank P. Carloni, A.V. Finkelstein and L. Tubiana for valuable discussions and suggestions.
PF is also a  member of the Interdisciplinary Laboratory for Computational 
Science (LISC), a joint venture between  University of Trento and 
Fondazione Bruno Kessler. T\v{S} is supported by the 
\emph{Provincia Autonoma di Trento}, through the AuroraScience project.
Simulations were performed on the AURORA supercomputer at the LISC.


\newpage

\section*{Figure Legends}

%

\newpage
\newpage

\newpage
\newpage

\newpage

\newpage
%
\newpage

\newpage

\end{document}